# Statistical Facts of Artificial Stock Market
## Comparison with Indonesian Empirical Data[*]

**Hokky Situngkir**[**]
(hokky@elka.ee.itb.ac.id)
Dept. Computational Sociology
Bandung Fe Institute

**Yohanes Surya**[***]
(yohaness@centrin.net.id)
Surya Research Intl.

**ABSTRACT**

The paper reports the construction of artificial stock market that emerges the similar statistical facts with real data in Indonesian stock market. We use the individual but dominant data, *i.e.*: PT TELKOM in hourly interval. The artificial stock market shows standard statistical facts, *e.g.*: volatility clustering, the excess kurtosis of the distribution of return, and the scaling properties with its breakdown in the crossover of Levy distribution to the Gaussian one. From this point, the artificial stock market will always be evaluated in order to have comprehension about market process in Indonesian stock market generally.

Stock market has been widely recognized as complex system with many interacting agents involve in the price formation. In return, the contemporary statistical mechanics incorporating in quantitative financial analysis has also employed the agent-based model *e.g.*: Ising model, to understand how the interacting agents shape the financial time series *e.g.*: price fluctuations. In the previous work, we have introduced some ways to cope with the need to understand stock market as complex system throughout the agent-based model (Situngkir & Surya, 2003b & 2004). We have reviewed some previous and important milestones in this endeavor, *e.g.*: some platforms of agent-based model (Farmer, 2001), the minority model (Challet, *et.al.*, 1999), the Santa Fe model (LeBaron, 2002), and the gate to the computational economics (Tesfatsion, 2002).

The paper presented here can be seen as a further advancement of agent-based model constructed in Situngkir & Surya (2004) to compare the price fluctuation produced by artificial market with the real data in Indonesian stock market (*i.e.*: hourly data January 2002 – September 2003 of individual index PT TELKOM). We see how the artificial stock market gives the similar price formation characters with the real data analyzed in Situngkir & Surya (2003a), i.e.:
- the volatility clustering
- the leptokurtic distribution, and
- the scaling properties.

**1. Model Overview**

The stock market is composed by heterogeneous interacting agents. In this sense, we can see that the price formation in the stock market is emerged by the heterogeneous strategies of investors or financial agents. Our artificial stock market is inspired by the formation of agents described in Castiglione (2001) and price-formation of market-making model (Farmer, 2001, elaborated also in Cont & Bouchaud, 2000), where there are about five types of agent, i.e.:
- *Fundamentalist strategy*, a strategy that always has tendency to hold a price at a certain value. Means it will sell if the price is higher than its fundamental value and vice versa buy for price lower than its fundamental value. In the running simulation, the change of fundamental value is randomized in certain interval or given externally.
- *Noisy strategy*. Choosing transaction actions of selling randomly with probability 0,5 but only buy if she feels save to sell, *i.e.*: find 2 other agents randomly that also sell.
- *Chartist strategy*, known as strategy for those who monitor market trend for certain history referred horizon – this method also known as moving average (MA). Agent sells if MA value:

$$\overline{m}_t(h) = \frac{1}{h}\sum_{t'=t-h}^{t-1} p_{t'} \qquad \ldots(1)$$

computed with h time horizon is larger than the price:

$$p_t^+(\delta) = p_t + p_t\delta \qquad \ldots(2)$$

where $\delta = (0,1)$ as input parameter. They will sell if the value of MA parameter is below the price: $p_t^-(\delta) = p_t - p_t\delta$. In the simulation, we have three types of this strategy differed by the horizon they use, *i.e.*: 30, 60, and 100 previous data.

Each agent occupies some agent-properties, *i.e.*:

- ✓ Choices of sell, in-active, or sell, represented as $x_t^{(i)} \in \{-1,0,1\}$
- ✓ Stock or capital that will be invested in stock market represented as $c_t^{(i)}$
- ✓ Number of stocks that become investment in stock market $n_t^{(i)}$. So that, in each iteration, the total asset of each agent: $k_t^{(i)} = n_t^{(i)} p_t + c_t^{(i)}$
- ✓ Influence strength: the agent's influence towards other agents on their decision to buy, hold, or sell, $r^{(i)} = [0,1], r^{(i)} \in \Re$.

As previously noted, every agent is allowed to buy or sell only one stock in every round. Agents are not allowed to do short-selling, since the sell or buy decisions must consider whether or not agent can afford with the transaction. An agent is forbidden to sell if she does not have any stock to sell, and in the other hand, agent cannot buy if she does not have enough money to do so.

As introduced in the previous work (Situngkir & Surya, 2004), the decision to sell, hold, or buy also consider the climate of the market, i.e.: the accumulated influence strength of all of the agents. Each agent affects and is affected by her surroundings on a variable of influence strength, say $s^{(i)}$. Each decision, $x_t^{(i)} \in \{-1,0,1\}$, is determined by agent's strategy - we normalize the value in the interval between –1,0,1. Therefore it can be seen as probability, i.e.:

$$P[x_t^{(i)} = x] = \frac{\sum_{j}^{x_t^{(j)}=x} s^{(j)}}{\sum_{j} s^{(j)}} \qquad \ldots(3)$$

with total possibility follows:

$$\sum_{x} P[x_t^{(i)} = x] = 1 \qquad \ldots(4)$$

The price emerged by the agent's interactions is calculated by the excess demand in each round, *i.e.*:

$$\Delta p = p(t+1) - p(t) = \frac{1}{\lambda}\left(\sum_{i}^{N} x_t^{(i)}\right) \quad \ldots(5)$$

where $\lambda$ is the market depth or liquidity, the excess demand needed to move the price by one unit. The market depth measures the sensitivity of price to fluctuations in excess demand (Cont & Bouchaud, 2000).

As a summary of the model overview, we can see table 1 showing the value of variables used in simulations.

**Table 1**
**Initial Simulation Configuration**

| Parameters | Value |
|---|---|
| Number of iteration | 10,000 |
| Number of agent (investor) | 200 |
| Formation *fundamentalist-chartist-noisy* | 42-109-49 |
| Chartist (h=30) – (h=60) – (h=100) | 46-33-30 |
| Stock owned by each agent | 10 |
| Money owned by each agent | IDR 20,000 |
| Market Depth ($1/\lambda$) | 10 |
| Basic price each stock | IDR 5,000 |

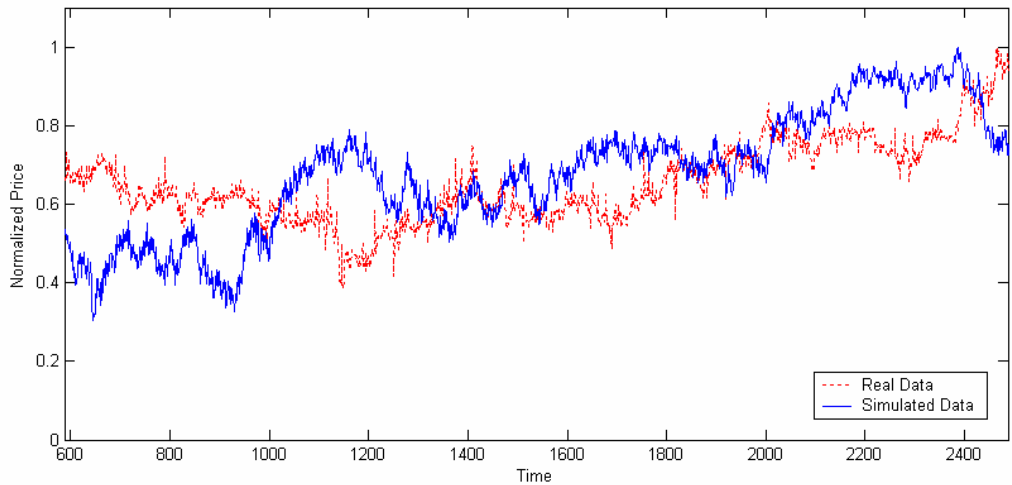

**Figure 1**
The simulation result compared to the real normalized hourly price data of a dominant individual index in Indonesia, PT TELKOM.

## 2. Simulation Results

We do several simulations in our artificial stock market in order to have some understanding points of what we discover in previous work on statistical properties of Indonesia stock market (Situngkir & Surya, 2003a, Hariadi & Surya, 2003). A pattern we want to analyze is the fact of volatility clustering, in which large changes tend to follow large changes, and small changes tend to follow small changes.

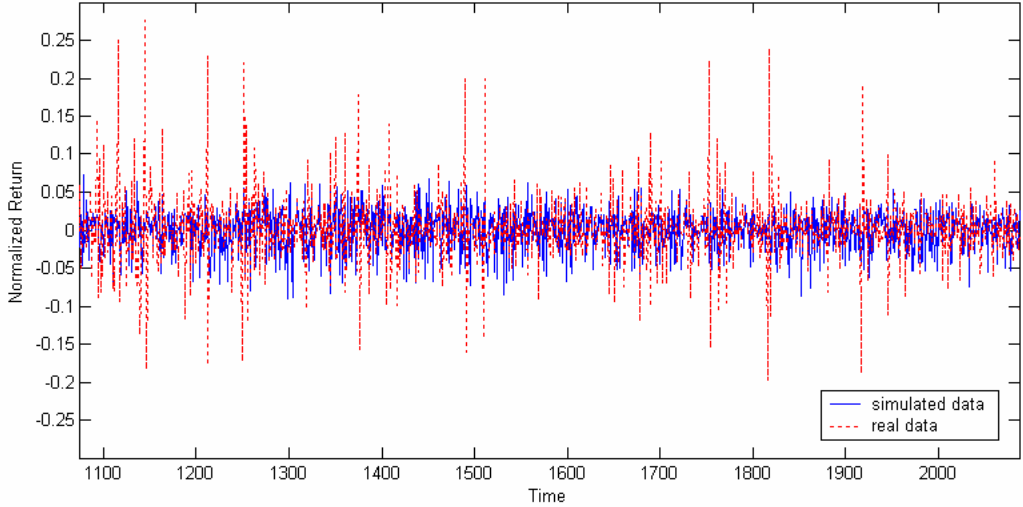

**Figure 2**
The return of simulated price fluctuations compared with the real data. The simulated data as the real one, exhibits similar pattern of volatility clustering.

The volatility clustering has been widely known as an important and interesting property of the financial time-series data. The cause of this property is certainly the interaction of between the heterogeneous agents; in our case: the fundamentalists, the chartists, and the noise traders. The decisions of any strategies will be different in the sense of expectations about future prices. Other important feature of our simulation is the boundedness of each agents one another on their final decisions; as noted above we apply the influence strength of any decisions (buy, hold, or sell) as the climate of the market. Henceforth, in certain time, a climate to sell, hold, or buy among agents becomes the trigger for the volatility clustering.

In advance, the volatility clustering has understood also impacts to the distribution of the financial data. The distribution of the price fluctuations (return) is less Gaussian with fat tails (leptokurtic) fitted with the truncated Levy distribution (Mantegna. & Stanley, 2000:60-67, Surya, *et.al.*, 2004) *i.e.*:

$$p(x) = \begin{cases} \xi L_{\alpha,0}(x) & -l \leq x \leq l \\ 0 & \text{otherwise} \end{cases} \quad \ldots(6)$$

where $\xi$ denotes the normalizing constant, $l$ the truncation parameter, and $L_{\alpha,0}$ the Levy distribution (whose coefficient $\alpha$ and $\beta = 0$). This is the form of distribution with finite variance and considering the Central Limit Theorem, which states that the sum of independent samples from any distribution with finite mean and variance converges to the Gaussian distribution as the sample size goes to infinity.

Figure 3 shows the distribution of the simulated return compared with the real data. The distribution of the return is leptokurtic, with fatter tail than Gaussian distribution.

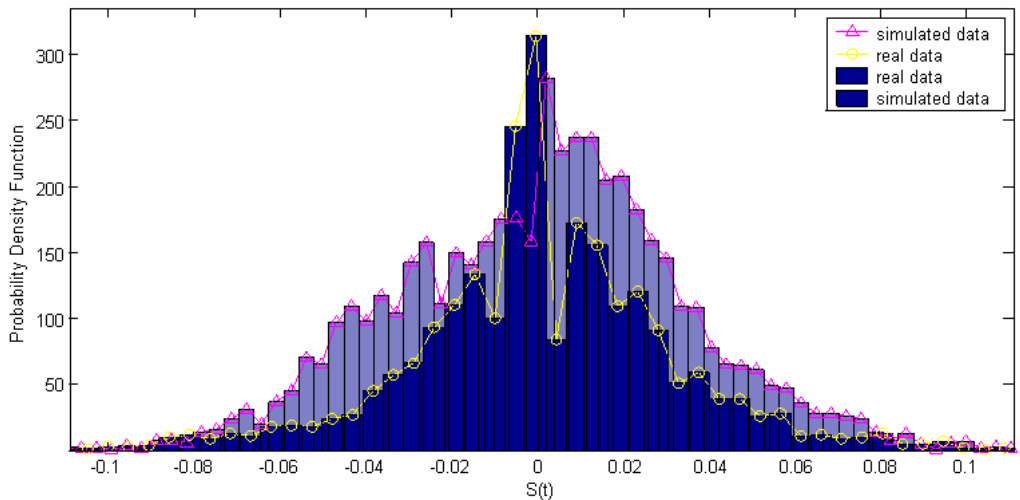

**Figure 3**
The probability density function of the simulated price fluctuation (return) compared to the real data showing the fat tail characteristics.

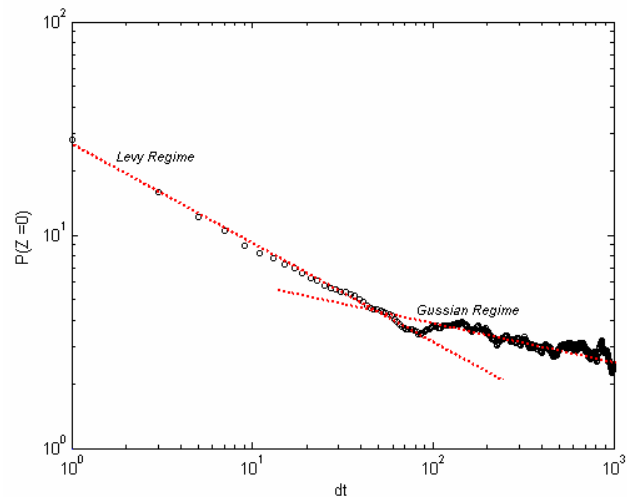

**Figure 4**
The probability of return to the origin of the simulated data shows the crossover from the Levy regime to the Gaussian regime as the consequence of the Central Limit Theorem.

Thus, we can see that the distribution of return of our simulated data also follows the Central Limit Theorem by fitting with the truncated Levy distribution. Let $\{X_i\}$ denotes the return of some financial data, the distribution of $\{X_i\}$ is estimated on the truncated Levy distribution and defined as $S_n:=X_1+X_2+\ldots+X_n$ and $Z_{dt}(t):=S_t-S_{t-dt}=X_t+X_{t-1}+\ldots+X_{t-dt+1}$. Thus, according to the Central Limit Theorem, $Z_{dt}$ will converge to a certain value of $dt$, say $dt=dt_x$ in which we have two distribution limit, *i.e.*: Levy and Gauss distribution. Mathematically,

$$p(S_{dt}) \approx \begin{cases} L_{\alpha,0}(S_{dt}), dt \ll dt_x \\ G(S_{dt}), dt \gg dt_x \end{cases} \quad \ldots(7)$$

where $G$ is the Gaussian Distribution, and the value of $dt$ presented as a parameter of the "distance" between the distribution of financial data to the normal distribution (Surya, *et.al.,* 2004:72-74).

Furthermore, this brings us to another important feature of empirical financial time series data, the scaling properties and its breakdown (Mantegna & Stanley, 2000). Roughly, as long as the distribution of return in the Levy regime, the data will have the scaling properties – but the scaling is breakdown when the crossover emerges on certain time-interval.

Thus, we have showed how the data of our artificial stock market has similar statistical properties with the real one, while the next step is finding important explanation of our stock market by comprehension on our structure of artificial stock market.

## 3. Discussions

We have constructed the artificial stock market that emerges the similar statistical facts with the real one for a certain individual but dominant index in Jakarta Stock Exchange, PT TELKOM. We have seen the volatility clustering and leptokurtic distribution of return of our simulation. The symptom of volatility clustering is seen as positive autocorrelation function and declining to reach zero. If the data shown in time series of $y_i$ with $i=1,2,3,\ldots$, thus the autocorrelation coefficient can be written as:

$$r_k = \frac{\sum_{i=1}^{n-k}(y_{i+k} - \bar{y}_{i+k})(y_i - \bar{y}_k)}{\sqrt{\sum_{i=1}^{n}(y_i - \bar{y}_i)^2 \times \sum_{i=1}^{n}(y_{i+k} - \bar{y}_{i+k})^2}} \quad \ldots(8)$$

where $r_k$ is the autocorrelation of $y_i$ and $y_{i+k}$. Autocorrelation of several samples of data forming distribution of around $k$ is commonly called *sampling* distribution autocorrelation. In Figure 5, we can find out that autocorrelation function of the real and simulated data presenting a similarity.

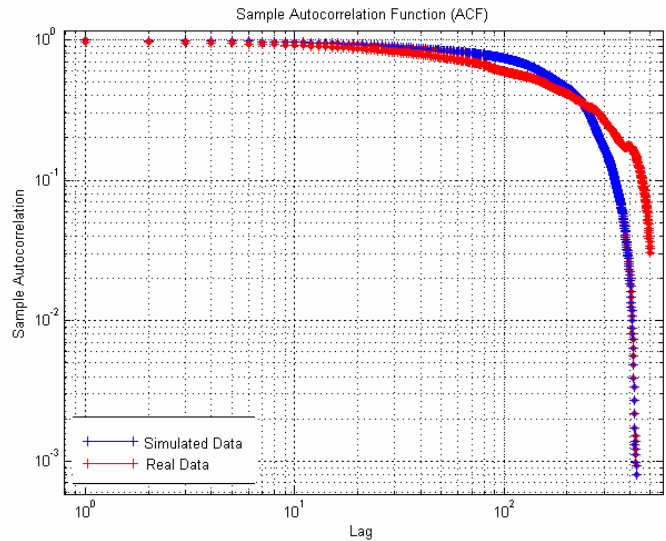

**Figure 5**
The sample autocorrelation function of the simulated data and the real one.

As noted above, we recognize that the volatility clustering is caused by the interaction among heterogeneous agents. The heterogeneity of our agents in the simulations is showed in table 1. It is obvious that the majority of agents' decisions depend on the trend of the price fluctuation, *i.e.*: the chartist. We can say (roughly) that most of the traders on PT TELKOM stocks follow the trend of the price fluctuation rather than try to keep the track of such fundamental values. However, further empirical researches on traders' strategies are important to verify this claim.

Other important note we can have as the result of the simulation is that the traders in Jakarta Stock Exchange are truly bounded by the climate of the market, since the tuning on the variable effect very sharply on the comparison to the real data. Once we give certain probability whether or not to follow the market climate, the simulated data become unrealistic. The simulation resulting figure 1 use the assumption that the traders are fully follow the market climate.

The data we use as comparison is the hourly individual index, henceforth it is important to proceed the model, *e.g.*: raise the heterogeneity of the agent's strategies or incorporating the price mechanism of the continuous market in which traders proposed the price and the stocks to be traded. This will be left in further research.

**4. Concluding Remarks**

We report the artificial stock market that emerges the similar statistical facts with real data. The data we use is the individual but dominant index, *i.e.*: hourly data of PT TELKOM in the time interval January 2002 up to September 2003. The artificial stock market shows standard statistical facts, *e.g.*: volatility clustering, the excess kurtosis of the distribution of return, and the scaling properties with its breakdown in the crossover of Levy distribution to the Gaussian one.

The advantage we can have by the simulation is the understanding of the interaction among traders and their composition of strategies in the Jakarta Stock Exchange. Practically, this can bring us a nice intuitive tool on comprehend the market mechanism in the stock market. Nonetheless, it should need much more further work, especially empirical one, in order to bring us more understanding of the market, *e.g.:* the rationality of the traders, strategies, and more about the decision making. In the other hand, we should construct more realistic market mechanism to have the long-term evolutionary structures of the market.

**Acknowledgement**

The authors thank the Surya Research Intl. for financial and data support, and some colleagues in BFI for important criticisms, especially Yun Hariadi for the technical discussions about multifractality. All faults remain the authors'.

**Work Cited:**

Bak. P., Paczuski, M., & Shubik, M. (1996). *Price Variations in a Stock Market with Many Agents*. Publikasi on-line di arXiv:cond-mat/9609144

Castiglione, Fillipo. (2001). *Microsimulation of Complex System Dynamics*. Inaugural Dissertation. Universität zu Köln

Challet, D., Marsili, M., & Zhang, Yi-Cheng. (1999). *Modeling Market Mechanism with Minority Game*. Pre-print: arxiv:cond-mat/9909265.

Cont, R. & Bouchaud, J.P. (2000). "Herd Behavior and Aggregate Fluctuations in Financial Market". *Macroeconomic Dynamics* 4:170-196.

Farmer, J. Doyne. (2001). *Toward Agent Based Models for Investment*. In *Benchmarks and Attribution Analysis*. Association for Investment and Management Research.

Hariadi, Y. & Surya, Y. (2003). *Multifraktal: Telkom, Indosat, & HMSP*. Working Paper WPT2003. Bandung Fe Institute.

LeBaron, Blake. (2002). *Building the Santa Fe Artificial Stock Market*. Woking Paper Brandeis University. URL: http://www.brandeis.edu/~blebaron

Mantegna, R. M. & Stanley, H.E. (2000). *An Introduction to Econophysics: Correlations and Complexity in Finance*. Cambridge University Press.

Situngkir, H. (2003). *Emerging the Emergence Sociology: The Philosophical Framework of Agent-Based Social Studies*, dalam *Journal of Social Complexity* Vol. 1(2):3-15, Bandung Fe Institute Press.

Situngkir, H. & Surya, Y. (2003a). *Platform Bangunan Multi-Agen dalam Analisis Keuangan: Gambaran Deskriptif Komputasi*. Working Paper WPS2003. Bandung Fe Institute.

Situngkir, H. & Surya, Y. (2003b). *Stylized Statistical Facts of Indonesian Financial Data: Empirical Study of Several Stock Indexes in Indonesia*. Working Paper WPU2003. Bandung Fe Institute. Pre-print: arxiv:cond-mat:0403465

Situngkir, H. & Surya, Y. (2004). *Agent-based Model Construction In Financial Economic System*. Working Paper WPA2004. Bandung Fe Institute. Pre-print: arxiv:nlin.AO/0403041

Surya, Y., Situngkir, H., Hariadi, Y., Suroso, R. (2004). *Aplikasi Fisika dalam Analisis Keuangan: Mekanika Statistika Interaksi Agen*. Bina Sumber Daya MIPA.

Tesfatison, Leigh. (2002). *Agent-Based Computational Economics: Growing Economics from the Bottom Up*. ISU Economic Working Paper No.1. Iowa State University.